\documentclass{article}
\usepackage{amsfonts}
\usepackage{amssymb}
\usepackage{amsmath}

\input{tcilatex}
\begin{document}

\title{Quantum Mechanics as Hamilton-Killing Flows on a Statistical Manifold%
\thanks{%
Presented at MaxEnt 2021, The 40th International Workshop on Bayesian
Inference and Maximum Entropy Methods in Science and Engineering, (July
5--9, 2021, TU Graz, Austria). }}
\author{Ariel Caticha \\
{\small Physics Department, University at Albany-SUNY, Albany, NY 12222, USA.%
}}
\date{ }
\maketitle

\begin{abstract}
The mathematical formalism of Quantum Mechanics is derived or
\textquotedblleft reconstructed\textquotedblright\ from more basic
considerations of probability theory and information geometry. The starting
point is the recognition that probabilities are central to QM: the formalism
of QM is derived as a particular kind of flow on a finite dimensional
statistical manifold --- a simplex. The cotangent bundle associated to the
simplex has a natural symplectic structure and it inherits its own natural
metric structure from the information geometry of the underlying simplex. We
seek flows that preserve (in the sense of vanishing Lie derivatives) both
the symplectic structure (a Hamilton flow) and the metric structure (a
Killing flow). The result is a formalism in which the Fubini-Study metric,
the linearity of the Schr\"{o}dinger equation, the emergence of a complex
numbers, Hilbert spaces, and the Born rule, are derived rather than
postulated.
\end{abstract}

\section{Introduction}

In the traditional approach to quantum mechanics (QM) the Hilbert space
plays a central, dominant role and probabilities are introduced, almost as
an afterthought, in order to provide the phenomenological link for handling
measurements. The uneasy coexistence of the Hilbert and the probabilistic
structures is reflected in the two separate modes of wave function
evolution: one is the linear and deterministic Schr\"{o}dinger evolution and
the other is the discontinuous and stochastic wave function collapse. It has
given rise to long-standing problems in the interpretation of the quantum
state itself \cite{Bell 1990}-\cite{Leifer 2014}.

These difficulties have motivated alternative approaches in which, rather
than postulating Hilbert spaces as the starting point, one recognizes that
probabilities play the dominant role; probabilities are not just an
accidental feature peculiar to quantum measurements. The goal there is to
derive or \textquotedblleft reconstruct\textquotedblright\ the mathematical
formalism of QM from more basic considerations of probability theory and
geometry. (See \emph{e.g.}, \cite{Nelson 1985}-\cite{Caticha 2019}) and
references therein.)

In the Entropic Dynamics (ED) approach the central object is the epistemic
configuration space, which is a statistical manifold --- a space in which
each point represents a probability distribution \cite{Caticha 2019}. In
this paper our goal is to discuss those special curves that could
potentially play the role of trajectories. What makes those curves special
is that they are adapted to the natural geometric structures on the
statistical manifold.

Two such structures are of central importance. The first is familiar from
statistics: all statistical manifolds have an intrinsic metric structure
given by the information metric \cite{Amari Nagaoka 2000}\cite{Caticha 2021}%
. The second is familiar from classical mechanics \cite{Arnold 1997}\cite%
{Souriau 1997}\cite{Schutz 1980}. Since we are interested in trajectories,
we are naturally led to consider the vectors that are tangent to such curves
and also the dual vectors, or covectors --- it is these objects that will be
used to represent the analogues of the velocities of probabilities and their
momenta. Vectors and covectors live in the so-called tangent and cotangent
spaces, respectively. It turns out that the statistical manifold plus all
its cotangent spaces is itself a manifold --- the cotangent bundle --- that
can be endowed with a second natural structure called \emph{symplectic}. In
mechanics the cotangent bundle is known as phase space, and the symplectic
transformations are known as canonical transformations.

There is an extensive literature on the symplectic and metric structures
inherent to QM. They have been discovered, independently rediscovered, and
extensively studied by many authors \cite{Hermann 1965}-\cite{Elze 2012}.
Their crucial insight is that those structures, being of purely geometrical
nature, are not just central to classical mechanics, they are central to
quantum mechanics too. Furthermore, the potential connection and relevance
of information geometry to various aspects of QM including its metric
structure\ has also been studied. \cite{Reginatto Hall 2011}-\cite{Caticha
2019}, \cite{Wootters 1981}-\cite{Molitor 2015}.

To characterize congruences of curves in the epistemic phase space --- or,
equivalently, the \emph{flows} on the cotangent bundle --- we must address
two problems. First, we must characterize the particular cotangent space and
the symplectic structure that is relevant to QM. This amounts to
establishing the correct conjugate momenta to be paired to the coordinates
which, in our case, are probabilities. In classical mechanics this pairing
is accomplished with the help of a Lagrangian $L(q,\dot{q})$ and the
prescription $p=\partial L/\partial \dot{q}$. Here we have no access to a
Lagrangian and a different criterion is adopted \cite{Caticha 2019}. The
second problem is to provide the cotangent bundle with a metric structure
that is compatible with the information metric of the underlying statistical
manifold. The issue is that cotangent bundles are not statistical manifolds
and the challenge is to identify the natural set of assumptions that lead to
the right metric structure.

We show that the flows that are relevant to quantum mechanics are those that
preserve (in the sense of vanishing Lie derivatives) both the symplectic
structure (a Hamilton flow) and the metric structure (a Killing flow). The
characterization of these Hamilton-Killing (HK) flows results in a formalism
that includes states described by rays; a geometry given by the Fubini-Study
metric, flows that obey a linear Schr\"{o}dinger equation, the emergence of
a complex structure, the Born rule, and Hilbert spaces. All these elements
are derived rather than postulated.

The present discussion includes two new developments. First, our focus will
be on isolating the essential geometrical aspects of the problem (a
discussion of the physical aspects is given in \cite{Caticha 2019}) and the
main ideas are presented in the simpler context of a finite-dimensional
manifold --- a simplex. Thus, what we derive here is the geometrical
framework that applies to a toy model --- an $n$-sided quantum die. And
second, the metric structure of the cotangent bundle is found by a new
argument involving the minimal assumption that the metric of phase space is
determined by the only metric structure at our disposal, namely the
information metric of the simplex.

Is this all there is to quantum mechanics? We conclude with a word of
caution: the framework developed here takes us a long way towards justifying
the mathematical formalism that underlies quantum mechanics, but it is only
a prelude to the true dynamics. The point is that not every HK curve is a
trajectory and not every parameter that labels points along a curve is time.
All changes of probabilities, including the changes we call dynamics, must
be compatible with the entropic and Bayesian rules that have been found to
be of universal applicability in inference. It is this additional
requirement that further restricts the HK flows to an entropic dynamics that
describes an evolution in a suitably constructed entropic concept of time 
\cite{Caticha 2010}\cite{Caticha 2019}.

This paper focuses on deriving the mathematical formalism of quantum
mechanics but the ED approach has been applied to a variety of other topics
in quantum theory. These include: the quantum measurement problem \cite%
{Johnson Caticha 2011}\cite{Vanslette Caticha 2016}; momentum and
uncertainty relations \cite{Nawaz Caticha 2011}\cite{Bartolomeo Caticha 2016}%
; the Bohmian limit \cite{Bartolomeo Caticha 2016}\cite{Bartolomeo Caticha
2015} and the classical limit \cite{Demme Caticha 2016}; extensions to
curved spaces \cite{Nawaz et al 2015}; to relativistic fields \cite{Ipek
Caticha 2014}\cite{Ipek Abedi Caticha 2018}\cite{Ipek Caticha 2020}; and the
ED of spin \cite{Caticha Carrara 2019}.

\section{Some background}

We deal with several distinct spaces. One is the \emph{ontic configuration
space} of microstates labelled by $i=1\ldots n$ --- these are the unknown
variables we are trying to predict. Another is the space of probability
distributions $\rho =(\rho ^{1}\ldots \rho ^{n})$ which is the \emph{%
epistemic configuration space} or, to use a shorter name, the \emph{%
e-configuration space}. This $(n-1)$-dimensional statistical manifold is a
simplex $\mathcal{S}$,%
\begin{equation}
\mathcal{S}=\left\{ \rho |\text{~}\rho ^{i}\geq
0\,;~\tsum\nolimits_{i=1}^{n}\rho ^{i}=1\right\} ~.
\end{equation}%
As coordinates for a generic point $\rho $ on $\mathcal{S}$ we shall use the
probabilities $\rho ^{i}$ themselves.

Given the manifold $\mathcal{S}$ we can construct two other special
manifolds that will turn out to be useful, the \emph{tangent bundle }$T%
\mathcal{S}$ and the \emph{cotangent bundle }$T^{\ast }\mathcal{S}$. These
are fiber bundles: the \emph{base manifold} is $\mathcal{S}$ and the fibers
at each point $\rho $ are respectively the tangent $T\mathcal{S}_{\rho }$
and cotangent $T^{\ast }\mathcal{S}_{\rho }$ spaces at $\rho $. The tangent
space at $\rho $, $T\mathcal{S}_{\rho }$, is the vector space composed of
all vectors that are tangent to curves through the point $\rho $. While this
space is obviously important (it is the space of \textquotedblleft
velocities\textquotedblright\ of probabilities) in what follows we will not
have much to say about it. Much more central to our discussion will be the
cotangent space at $\rho $, $T^{\ast }\mathcal{S}_{\rho }$, is the vector
space of all covectors at $\rho $ (which is the space of the momenta
conjugate to those probabilities).

As already mentioned, the reason we care about vectors and covectors is that
these are the objects that will eventually be used to represent velocities
and momenta. The cotangent bundle $T^{\ast }\mathcal{S}$, will play the
central role of the \emph{epistemic phase space}, or \emph{e-phase space.}

A point $X\in T^{\ast }\mathcal{S}$ will be represented as $X=\left( \rho
,\pi \right) $, where $\rho =(\rho ^{1}\ldots \rho ^{n})$ are coordinates on
the base manifold $\mathcal{S}$ and $\pi =(\pi _{1}\ldots \pi _{n})$ are
some generic coordinates on the cotangent space $T^{\ast }\mathcal{S}_{\rho
} $ at $\rho $. Curves on $T^{\ast }\mathcal{S}$ allow us to define vectors
on the tangent spaces $T(T^{\ast }\mathcal{S})_{X}$. Let $X=X(\lambda )$ be
a curve parametrized by $\lambda $, then the vector $\bar{V}$ tangent to the
curve at $X=(\rho ,\pi )$ has components $d\rho ^{i}/d\lambda $ and $d\pi
_{i}/d\lambda $, and is written 
\begin{equation}
\bar{V}=\frac{d}{d\lambda }=\frac{d\rho ^{i}}{d\lambda }\bar{\rho}_{i}+\frac{%
d\pi _{i}}{d\lambda }\bar{\pi}^{i}=\frac{d\rho ^{i}}{d\lambda }\frac{%
\partial }{\partial \rho ^{i}}+\frac{d\pi _{i}}{d\lambda }\frac{\partial }{%
\partial \pi _{i}}~,  \label{s vector}
\end{equation}%
where $\bar{\rho}_{i}$ and $\bar{\pi}^{i}$ are the basis vectors, the index $%
i=1\ldots n$ is summed over, and we adopt the standard notation in
differential geometry, $\bar{\rho}_{i}=\partial /\partial \rho ^{i}$ and $%
\bar{\pi}^{i}=\partial /\partial \pi _{i}$. The directional derivative of a
function $F(X)$ along the curve $X(\lambda )$ is 
\begin{equation}
\frac{dF}{d\lambda }=\frac{\partial F}{\partial \rho ^{i}}\frac{d\rho ^{i}}{%
d\lambda }+\frac{\partial F}{\partial \pi _{i}}\frac{d\pi _{i}}{d\lambda }%
\overset{\text{def}}{=}\mathsf{\tilde{\nabla}}F[\bar{V}]~,
\label{s dir deriv a}
\end{equation}%
where $\mathsf{\tilde{\nabla}}$ is the gradient in $T^{\ast }\mathcal{S}$,
that is, the gradient of a generic function $F(X)=F(\rho ,\pi )$ is 
\begin{equation}
\mathsf{\tilde{\nabla}}F=\,\frac{\partial F}{\partial \rho ^{i}}\mathsf{%
\tilde{\nabla}}\rho ^{i}+\frac{\partial F}{\partial \pi _{i}}\mathsf{\tilde{%
\nabla}}\pi _{i}~,  \label{s gradient}
\end{equation}%
where $\mathsf{\tilde{\nabla}}\rho ^{i}$ and $\mathsf{\tilde{\nabla}}\pi
_{i} $ are the basis covectors. The tilde `$\symbol{126}$' serves to
distinguish the gradient $\tilde{\nabla}$ on the bundle $T^{\ast }\mathcal{S}
$ from the gradient $\nabla $ on the simplex $\mathcal{S}$.

Here, unfortunately, we encounter a technical difficulty due to the fact
that the space $\mathcal{S}$ is constrained to normalized probabilities so
that the coordinates $\rho ^{i}$ cannot be varied independently. This
problem is handled, without loss of generality, by embedding the $(n-1)$%
-dimensional manifold $\mathcal{S}$ into a manifold of one dimension higher,
the so-called positive-cone, denoted $\mathcal{S}^{+}$, where the
coordinates $\rho ^{i}$ are unconstrained.

To simplify the notation further a point $X=(\rho ,\pi )$ in the $2n$%
-dimensional $T^{\ast }\mathcal{S}^{+}$ will be labelled by its coordinates $%
X^{\alpha i}=(X^{1i},X^{2i})=(\rho ^{i},\pi _{i})$, where $\alpha i$ is a
composite index. The first index $\alpha $ (chosen from the beginning of the
Greek alphabet) takes two values, $\alpha =1,2$. Since $\alpha $ keeps track
of whether $i$ is an upper $\rho ^{i}$ index ($\alpha =1$) or a lower $\pi
_{i}$ index ($\alpha =2$), from now on we can set $\rho _{i}=\rho ^{i}$.
Then eqs.(\ref{s vector}) and (\ref{s gradient}) are written as 
\begin{equation}
\bar{V}=\frac{d}{d\lambda }=V^{\alpha i}\frac{\partial }{\partial X^{\alpha
i}}\quad \text{and}\quad \mathsf{\tilde{\nabla}}F=\,\frac{\partial F}{%
\partial X^{\alpha i}}\mathsf{\tilde{\nabla}}X^{\alpha i}~.  \label{s grad a}
\end{equation}%
The repeated indices indicate a double summation over $\alpha $ and $i$. The
action of the basis covectors $\mathsf{\tilde{\nabla}}X^{\alpha i}$ on the
basis vectors, $\partial /\partial X^{\beta j}=\partial _{\beta j}$ is given
by 
\begin{equation}
\mathsf{\tilde{\nabla}}X^{\alpha i}[\partial _{\beta j}]=\frac{\partial
X^{\alpha i}}{\partial X^{\beta j}}=\delta _{\beta j}^{\alpha i}\quad \text{%
so that}\quad \mathsf{\tilde{\nabla}}F[\bar{V}]=\frac{\partial F}{\partial
X^{\alpha i}}V^{\alpha i}=\frac{dF}{d\lambda }  \label{s grad b}
\end{equation}%
is the directional derivative of $F$ along the vector $\bar{V}$.

\section{Hamiltonian flows}

Just as a manifold can be supplied with a symmetric bilinear form, the
metric tensor, which gives it the fairly rigid structure described as its 
\emph{metric} \emph{geometry}, cotangent bundles can be supplied with an 
\emph{antisymmetric} bilinear form, the \emph{symplectic form}, which gives
them the somewhat floppier structure called \emph{symplectic geometry} \cite%
{Arnold 1997}\cite{Souriau 1997}\cite{Schutz 1980}.

A vector field $\bar{V}(X)$ defines a space-filling congruence of curves $%
X^{i}=X^{i}(\lambda )$ that are tangent to the field $\bar{V}(X)$ at every
point $X$. We seek those special congruences or \emph{flows} that reflect
the symplectic geometry.

\subsubsection*{\textbf{The symplectic form}}

\label{Symplectic Form}

Once local coordinates $(\rho ^{i},\pi _{i})$ on $T^{\ast }\mathcal{S}^{+}$
have been established there is a natural choice of symplectic form 
\begin{equation}
\Omega =\,\tilde{\nabla}\rho ^{i}\otimes \tilde{\nabla}\pi _{i}-\tilde{\nabla%
}\pi _{i}\otimes \tilde{\nabla}\rho ^{i}~.  \label{sympl form a}
\end{equation}%
The question of how to choose those local coordinates, which are Darboux
coordinates for the cotangent bundle, remains open. The answer is not to be
found in mathematics but in physics. In classical mechanics the criterion
for choosing a canonical momentum is provided by a Lagrangian, but here we
do not have a Lagrangian. An alternative criterion more closely taylored to
the framework presented here is provided by Entropic Dynamics \cite{Caticha
2019}. From now on we shall assume that the correct $\pi _{i}$ coordinates
have been identified.

The action of $\Omega \lbrack \cdot ,\cdot ]$ on two vectors $\bar{V}%
=d/d\lambda $ and $\bar{U}=d/d\mu $ is obtained using (\ref{s grad b}), 
\begin{equation}
\tilde{\nabla}\rho ^{i}(\bar{V})=V^{1i}\quad \text{and}\quad \tilde{\nabla}%
\pi _{i}(\bar{V})=V^{2i}~.
\end{equation}%
The result is 
\begin{equation}
\Omega \lbrack \bar{V},\bar{U}]=V^{1i}U^{2i}-V^{2i}U^{1i}=\Omega _{\alpha
i,\beta j}V^{\alpha i}U^{\beta j}\quad \text{where}\quad \Omega _{\alpha
i,\beta j}=%
\begin{bmatrix}
0 & 1 \\ 
-1 & 0%
\end{bmatrix}%
\delta _{ij}~.~  \label{sympl form b}
\end{equation}

\subsubsection*{\textbf{Hamilton's equations and Poisson brackets}}

Next we derive the $2n$-dimensional $T^{\ast }\mathcal{S}^{+}$ analogues of
results that are standard in classical mechanics \cite{Arnold 1997}\cite%
{Souriau 1997}\cite{Schutz 1980}. We seek those vector fields $\bar{V}(X)$
that generate flows (the congruence of integral curves) that preserve the
symplectic structure in the sense that 
\begin{equation}
\pounds _{V}\Omega =0~,
\end{equation}%
where the Lie derivative \cite{Schutz 1980} is 
\begin{equation}
(\pounds _{V}\Omega )_{\alpha i,\beta j}=V^{\gamma k}\partial _{\gamma
k}\Omega _{\alpha i,\beta j}+\Omega _{\gamma k,\beta j}\partial _{\alpha
i}V^{\gamma k}+\Omega _{\alpha i,\gamma k}\partial _{\beta j}V^{\gamma k}~.
\end{equation}%
Since by eq.(\ref{sympl form b}) the components $\Omega _{\alpha i,\beta j}$
are constant, $\partial _{\gamma k}\Omega _{\alpha i,\beta j}=0$, we can
rewrite $\pounds _{V}\Omega $ as 
\begin{equation}
(\pounds _{V}\Omega )_{\alpha i,\beta j}=\partial _{\alpha i}(\Omega
_{\gamma k,\beta j}V^{\gamma k})-\partial _{\beta j}(\Omega _{\gamma
k,\alpha i}V^{\gamma k})~,
\end{equation}%
which is the exterior derivative (roughly, the curl) of the covector $\Omega
_{\gamma k,\alpha i}V^{\gamma k}$. By Poincare's lemma, requiring $\pounds %
_{V}\Omega =0$ (a vanishing curl) implies that $\Omega _{\gamma k,\alpha
i}V^{\gamma k}$ is the gradient of a scalar function, which we will denote $%
\tilde{V}(X)$,%
\begin{equation}
\Omega _{\gamma k,\alpha i}V^{\gamma k}=\partial _{\alpha i}\tilde{V}\quad 
\text{or}\quad \Omega (\bar{V},\cdot )=\tilde{\nabla}\tilde{V}(\cdot )~.
\label{grad V}
\end{equation}%
In the opposite direction we can easily check that (\ref{grad V}) implies $%
\pounds _{V}\Omega =0$. Using (\ref{sympl form b}), eq.(\ref{grad V}) is
more explicitly written as%
\begin{equation}
\frac{d\rho ^{i}}{d\lambda }\tilde{\nabla}\pi _{i}-\frac{d\pi _{i}}{d\lambda 
}\tilde{\nabla}\rho ^{i}=\frac{\partial \tilde{V}}{\partial \rho ^{i}}\tilde{%
\nabla}\rho ^{i}+\frac{\partial \tilde{V}}{\partial \pi _{i}}\tilde{\nabla}%
\pi _{i}~,
\end{equation}%
or 
\begin{equation}
\frac{d\rho ^{i}}{d\lambda }=\frac{\partial \tilde{V}}{\partial \pi _{i}}%
\quad \text{and}\quad \frac{d\pi _{i}}{d\lambda }=-\frac{\partial \tilde{V}}{%
\partial \rho ^{i}}~,  \label{Hamiltonian flow a}
\end{equation}%
which we recognize as Hamilton's equations for a Hamiltonian function $%
\tilde{V}$. This justifies calling $\bar{V}$ the \emph{Hamiltonian vector
field} associated to the \emph{Hamiltonian function} $\tilde{V}$. In words:
the flows that preserve the symplectic structure, $\pounds _{V}\Omega =0$,
are generated by Hamiltonian vector fields $\bar{V}$ associated to
Hamiltonian functions $\tilde{V}$.

From (\ref{sympl form b}) and (\ref{Hamiltonian flow a}) the action of the
symplectic form $\Omega $ on two Hamiltonian vector fields $\bar{V}%
=d/d\lambda $ and $\bar{U}=d/d\mu $ generated respectively by $\tilde{V}$
and $\tilde{U}$ is%
\begin{equation}
\Omega \lbrack \bar{V},\bar{U}]=\frac{d\rho ^{i}}{d\lambda }\frac{d\pi _{i}}{%
d\mu }-\frac{d\pi _{i}}{d\lambda }\frac{d\rho ^{i}}{d\mu }=\frac{\partial 
\tilde{V}}{\partial \rho ^{i}}\frac{\delta \tilde{U}}{\delta \pi _{i}}-\frac{%
\partial \tilde{V}}{\partial \pi _{i}}\frac{\partial \tilde{U}}{\partial
\rho ^{i}}\overset{\text{def}}{=}\{\tilde{V},\tilde{U}\}~,  \label{PB a}
\end{equation}%
where, on the right hand side, we have introduced the Poisson bracket
notation. In words, the action of $\Omega $ on two Hamiltonian vector fields
is the Poisson bracket of the associated Hamiltonian functions. We can also
check that the derivative of an arbitrary function $F(X)$ along the vector
field $\bar{V}=d/d\lambda $, is 
\begin{equation}
\frac{dF}{d\lambda }=\{F,\tilde{V}\}~.  \label{PB b}
\end{equation}

Thus, \emph{the Hamiltonian formalism that is so familiar in physics emerges
from purely geometrical considerations}. It might be desirable to adopt a
more suggestive notation: instead of $(\tilde{V},\lambda )$ let us write $(%
\tilde{H},\tau )$. Then the flow generated by a Hamiltonian function $\tilde{%
H}$ and parametrized by \textquotedblleft time\textquotedblright\ $\tau $ is
given by Hamilton's equations in the standard form, 
\begin{equation}
\frac{d\rho ^{i}}{d\tau }=\frac{\partial \tilde{H}}{\partial \pi _{i}}\quad 
\text{and}\quad \frac{d\pi _{i}}{d\tau }=-\frac{\partial \tilde{H}}{\partial
\rho ^{i}}~,  \label{Hamiltonian flow c}
\end{equation}%
and the $\tau $ evolution of any well-behaved function $f(X)$ is given by 
\begin{equation}
\frac{df}{d\tau }=\bar{H}(f)=\{f,\tilde{H}\}\quad \text{with}\quad \bar{H}=%
\frac{\partial \tilde{H}}{\partial \pi _{i}}\frac{\partial }{\partial \rho
^{i}}-\frac{\partial \tilde{H}}{\partial \rho ^{i}}\frac{\partial }{\partial
\pi _{i}}~.  \label{Hamiltonian flow d}
\end{equation}%
The difference with classical mechanics is that here the degrees of freedom
are probabilities and not ontic variables such as, for example, the
positions of particles.

\subsubsection*{\textbf{The normalization constraint}}

Since our actual interest is not in flows on $T^{\ast }\mathcal{S}^{+}$ but
on the bundle $T^{\ast }\mathcal{S}$ of normalized probabilities we shall
constrain to flows that preserve the normalization of probabilities. Let%
\begin{equation}
\left\vert \rho \right\vert \overset{\text{def}}{=}\tsum\limits_{i=1}^{n}%
\rho ^{i}\quad \text{and}\quad \tilde{N}\overset{\text{def}}{=}1-\left\vert
\rho \right\vert ~.
\end{equation}%
We seek those special Hamiltonians $\tilde{H}$ such that the initial
condition $\tilde{N}=0$ is preserved by the flow, that is, 
\begin{equation}
\partial _{\tau }\tilde{N}=\{\tilde{N},\tilde{H}\}=0\quad \text{or}\quad
\tsum\nolimits_{i}\frac{\partial \tilde{H}}{\partial \pi _{i}}%
=\tsum\nolimits_{i}\frac{d\rho ^{i}}{d\tau }=0~.  \label{N conservation}
\end{equation}%
Since the probabilities $\rho ^{i}$ must remain positive we shall further
require that $d\rho ^{i}/d\tau \geq 0$ when$\ \rho ^{i}=0$. \ 

We can also consider the Hamiltonian flow generated by $\tilde{N}$ and
parametrized by $\nu $. From eq.(\ref{Hamiltonian flow a}) the corresponding
Hamiltonian vector field $\bar{N}$ is given by 
\begin{equation}
\bar{N}=N^{\alpha i}\frac{\partial }{\partial X^{\alpha i}}\quad \text{with}%
\quad N^{\alpha i}=\frac{dX^{\alpha i}}{d\nu }=\{X^{\alpha i},\tilde{N}\}~,
\label{N vector a}
\end{equation}%
or, more explicitly, 
\begin{equation}
N^{1i}=\frac{d\rho ^{i}}{d\nu }=0\quad \text{and}\quad N^{2i}=\frac{d\pi _{i}%
}{d\nu }=1~.  \label{N vector b}
\end{equation}%
The integral curves generated by $\tilde{N}$ are found by integrating (\ref%
{N vector b}). The result is 
\begin{equation}
\rho ^{i}(\nu )=\rho ^{i}(0)\quad \text{and}\quad \pi _{i}(\nu )=\pi
_{i}(0)+\nu ~,  \label{momentum shift}
\end{equation}%
which amounts to shifting all momenta by the $i$-independent parameter $\nu $%
. We can also see that if $\tilde{N}$ is conserved along $\bar{H}$, then $%
\tilde{H}$ is conserved along $\bar{N}$, 
\begin{equation}
\frac{d\tilde{H}}{d\nu }=\{\tilde{H},\tilde{N}\}=0~,
\label{global gauge sym}
\end{equation}%
which implies that the conserved quantity $\tilde{N}$ is the generator of a
symmetry transformation.

The phase space of interest is $T^{\ast }\mathcal{S}$ but the description is
simplified by using the unnormalized coordinates $\rho $ of the larger
embedding space $T^{\ast }\mathcal{S}^{+}$. The introduction of one
superfluous $\rho $ coordinate forces us to also introduce one superfluous $%
\pi $ momentum. We eliminate the extra coordinate by imposing the constraint 
$\tilde{N}=0$. We eliminate the extra momentum by declaring it unphysical:
the shifted point $\left( \rho ^{\prime },\pi ^{\prime }\right) =\left( \rho
,\pi +\nu \right) $ is declared to be equivalent to $\left( \rho ,\pi
\right) $. This equivalence is described as a global \textquotedblleft
gauge\textquotedblright\ symmetry which, as we shall later see, is the
reason why quantum mechanical states are represented by rays rather than
vectors in a Hilbert space.

\section{The information geometry of e-phase space}

Our next goal is to extend the metric of the simplex $\mathcal{S}$ --- given
by information geometry --- to the full e-phase space, $T^{\ast }\mathcal{S}$%
. The extension can be carried out in many ways \cite{Reginatto Hall 2011}%
\cite{Reginatto Hall 2012}\cite{Caticha 2019}\cite{Caticha 2017}. The virtue
of the derivation below is that the number of input assumptions is kept to a
minimum.

\subsubsection*{\textbf{The metric on the embedding e-phase space }$T^{\ast
}S^{+}$}

First we assign a metric to the embedding bundle $T^{\ast }\mathcal{S}^{+}$
and then we shall consider the metric it induces on $T^{\ast }\mathcal{S}$.
The metric of the space $\mathcal{S}^{+}$ of unnormalized probabilities \cite%
{Caticha 2021}\cite{Campbell 1986} is 
\begin{equation}
\delta \ell ^{2}=g_{ij}\delta \rho ^{i}\delta \rho ^{j}\quad \text{with}%
\quad g_{ij}=A(|\rho |)n_{i}n_{j}+\frac{B(|\rho |)}{2\rho ^{i}}\delta _{ij}~,
\label{info metric ss c}
\end{equation}%
where $n$ is a covector with components $n_{i}=1$ for all $i=1\ldots n$, and 
$A(|\rho |)$ and $B(|\rho |)$ are smooth scalar functions of $|\rho |{}=\sum
\rho ^{i}$. Since \emph{the only tensor at our disposal} is $g_{ij}$ the
length element of $T^{\ast }\mathcal{S}^{+}$ must be of the form 
\begin{equation}
\delta \tilde{\ell}^{2}=\alpha g_{ij}\delta \rho ^{i}\delta \rho ^{j}+\beta
g_{i}^{j}\delta \rho ^{i}\delta \pi _{j}+\gamma g^{ij}\delta \pi _{i}\delta
\pi _{j}~,  \label{G+ abc}
\end{equation}%
where $\alpha $, $\beta $, and $\gamma $ are constants. Since $\delta \rho
^{i}$ and $\delta \pi _{i}$ are vectors and covectors the requirement that $%
\delta \tilde{\ell}^{2}$ induce the same magnitudes $g_{ij}\delta \rho
^{i}\delta \rho ^{j}$ on $T\mathcal{S}_{\rho }^{+}$ and $g^{ij}\delta \pi
_{i}\delta \pi _{j}$ on $T^{\ast }\mathcal{S}_{\rho }^{+}$ as given by
information geometry implies that $\alpha =\gamma =1$.  To fix $\beta $
consider a curve $\rho =\rho (\tau )$ and $\pi =\pi (\tau )$ on $T^{\ast }%
\mathcal{S}^{+}$ and its flow-reversed or $\tau $-reversed curve given by $%
\rho ^{\prime }(\tau )=\rho (-\tau )$ and $\pi ^{\prime }(\tau )=-\pi (-\tau
)$. We shall require that the speed $(d\tilde{\ell}/d\tau )^{2}$ remain
invariant under flow-reversal. Since under flow-reversal the mixed $\rho \pi 
$ terms in (\ref{G+ abc}) change sign, it follows that invariance implies
that $\beta =0$. We emphasize that imposing that e-phase space be symmetric
under flow-reversal does not amount to imposing \emph{time}-reversal
invariance; time-reversal violations might still be caused by interaction
terms in the Hamiltonian. The resulting line element, \emph{which has been
designed to be fully determined by information geometry}, takes the form, 
\begin{equation}
\delta \tilde{\ell}^{2}=G_{\alpha i,\beta j}\delta X^{\alpha i}\delta
X^{\beta j}=g_{ij}\delta \rho ^{i}\delta \rho ^{j}+g^{ij}\delta \pi
_{i}\delta \pi _{j}~.  \label{G+ a}
\end{equation}

\subsubsection*{\textbf{A complex structure for }$T^{\ast }S^{+}$}

The metric tensor $G$ and its inverse $G^{-1}$ can be used to lower and
raise indices. In particular, with $G^{-1}$ we can raise the first index of
the symplectic form $\Omega _{\alpha i,\beta j}$ in eq.(\ref{sympl form b}) 
\begin{equation}
G^{\alpha i,\gamma k}\Omega _{\gamma k,\beta j}\overset{\text{def}}{=}%
-J^{\alpha i}{}_{\beta j}~.  \label{J+ a}
\end{equation}%
The tensor $J$ has several important properties. These are most easily
derived by writing $G$ and $\Omega $ in block matrix form, 
\begin{equation}
G^{-1}=%
\begin{bmatrix}
g^{-1} & 0 \\ 
0 & g%
\end{bmatrix}%
\,,~\Omega =%
\begin{bmatrix}
0 & 1 \\ 
-1 & 0%
\end{bmatrix}%
~,~J=%
\begin{bmatrix}
0 & -g^{-1} \\ 
g & 0%
\end{bmatrix}%
~.  \label{G+ c}
\end{equation}%
We can immediately check that $JJ=-\mathbf{1}$ which shows that $J$ is a
square root of the negative identity matrix. Thus $J$ endows $T^{\ast }%
\mathcal{S}^{+}$ with a complex structure. \emph{To summarize: in addition
to the symplectic }$\Omega $\emph{\ and metric }$G$\emph{\ structures the
cotangent bundle }$T^{\ast }S^{+}$\emph{\ is also endowed with a complex
structure }$J$\emph{.} Such highly structured spaces are generically known
as K\"{a}hler manifolds. Here we deal with a special K\"{a}hler manifold
where the space of $\rho $s is a statistical manifold and the spaces of $\pi 
$s are flat cotangent spaces. But ultimately the geometry of $T^{\ast }%
\mathcal{S}^{+}$ is only of marginal interest; what matters is the geometry
it induces on the e-phase space $T^{\ast }\mathcal{S}$ of normalized
probabilities to which we turn next.

\subsubsection*{\textbf{The metric induced on the e-phase space }$T^{\ast }S$%
}

We saw that the e-phase space $T^{\ast }\mathcal{S}$ can be obtained from
the space $T^{\ast }\mathcal{S}^{+}$ by the restriction $|\rho |\,=1$ and by
identifying the gauge equivalent points $(\rho ^{i},\pi _{i})$ and $(\rho
^{i},\pi _{i}+n_{i}\nu )$. Consider two neighboring\ points $(\rho ^{i},\pi
_{i})$ and $(\rho ^{\prime i},\pi _{i}^{\prime })$ with $|\rho |\,=|\rho
^{\prime }|\,=1$. The metric induced on $T^{\ast }\mathcal{S}$ will be
defined as the shortest $T^{\ast }\mathcal{S}^{+}$ distance between $(\rho
^{i},\pi _{i})$ and points on the ray defined by $(\rho ^{\prime i},\pi
_{i}^{\prime })$. Since the $T^{\ast }\mathcal{S}^{+}$ distance between $%
(\rho ^{i},\pi _{i})$ and $(\rho ^{i}+\delta \rho ^{i},\pi _{i}+\delta \pi
_{i}+n_{i}\nu )$ is 
\begin{equation}
\delta \tilde{\ell}^{2}(\nu )=g_{ij}\delta \rho ^{i}\delta \rho
^{j}+g^{ij}(\delta \pi _{i}+n_{i}\nu )(\delta \pi _{j}+n_{i}\nu )~,
\label{TS metric a}
\end{equation}%
the metric on $T^{\ast }\mathcal{S}$ will be defined by $\delta \tilde{s}%
^{2}=\min_{\nu }\delta \tilde{\ell}^{2}$. Imposing $|\delta \rho |\,=0$, the
value of $\nu $ that minimizes (\ref{TS metric a}) is $\nu =-\langle \delta
\pi \rangle =-\tsum\nolimits_{i}\rho ^{i}\delta \pi _{i}$. Therefore, the
metric on $T^{\ast }\mathcal{S}$, which measures the distance between
neighboring rays, is 
\begin{equation}
\delta \tilde{s}^{2}=\tsum\limits_{i=1}^{n}\left[ \frac{B(1)}{2\rho ^{i}}%
(\delta \rho ^{i})^{2}+\frac{2\rho ^{i}}{B(1)}(\delta \pi _{i}-\langle
\delta \pi \rangle )^{2}\right] ~.  \label{TP metric c}
\end{equation}%
From now on we shall set $B(1)=1$, which only amounts to a choice of units
and has no effect on our results. (In \cite{Caticha 2019} we chose $%
B(1)=\hbar $.)

Although the metric (\ref{TP metric c}) is expressed in a notation that may
be unfamiliar, it turns out to be equivalent to the well-known \emph{%
Fubini-Study metric}. Thus, the recognition that the e-phase space is the
cotangent bundle of a statistical manifold has led us to a novel derivation
based on information geometry.

An important feature of the $T^{\ast }\mathcal{S}$ metric (\ref{TP metric c}%
) is that, except for the irrelevant constant $B(1)$, it has turned out to
be independent of the particular choices of the \emph{functions} $A(|\rho |)$
and $B(|\rho |)$ (see eq.(\ref{info metric ss c})) that define the
geometries of the embedding spaces $\mathcal{S}^{+}$ and $T^{\ast }\mathcal{S%
}^{+}$. Therefore, without any loss of generality we can simplify the
analysis considerably by choosing $A(|\rho |)=0$ and $B(|\rho |)=1$ which
gives the embedding spaces the simplest possible geometries, namely, they
are flat. With this choice the $T^{\ast }\mathcal{S}^{+}$ metric, eq.(\ref%
{G+ a}), becomes 
\begin{equation}
\delta \tilde{\ell}^{2}=\tsum\limits_{i=1}^{n}\left[ \frac{1}{2\rho ^{i}}%
\delta \rho _{i}^{2}+2\rho ^{i}\delta \pi _{i}^{2}\right] =G_{\alpha i,\beta
j}\delta X^{\alpha i}\delta X^{\beta j}\quad \text{with}\quad G_{\alpha
i,\beta j}=%
\begin{bmatrix}
\delta _{ij}/2\rho _{i} & 0 \\ 
0 & 2\rho _{i}\delta _{ij}%
\end{bmatrix}
\label{ePS a}
\end{equation}%
and the tensor $J$, eq.(\ref{G+ c}), which defines the complex structure,
becomes 
\begin{equation}
J^{\alpha i}{}_{\beta j}=-G^{\alpha i,\gamma k}\Omega _{\gamma k,\beta j}=%
\begin{bmatrix}
0 & -2\rho _{i}\delta _{ij} \\ 
\delta _{ij}/2\rho _{i} & 0%
\end{bmatrix}%
~.  \label{J tensor}
\end{equation}

\subsubsection*{\textbf{Refining the choice of cotangent space: complex
coordinates}}

Having endowed the e-phase spaces $T^{\ast }\mathcal{S}^{+}$ and $T^{\ast }%
\mathcal{S}$ with both metric and complex structures we can now revisit and
refine our choice of cotangent spaces. So far we had assumed the cotangent
space $T^{\ast }\mathcal{S}_{\rho }^{+}$ at $\rho $ to be the flat Euclidean 
$n$-dimensional space $R^{n}$. It turns out that the cotangent space that is
relevant to quantum mechanics requires a further restriction. To see what
this is we use the fact that $T^{\ast }\mathcal{S}^{+}$ is endowed with a
complex structure which suggests a coordinate transformation from $(\rho
,\pi )$ to complex coordinates $(\psi ,i\psi ^{\ast })$, 
\begin{equation}
\psi _{i}=\rho _{i}^{1/2}e^{i\pi _{i}}\quad \text{and}\quad i\psi _{i}^{\ast
}=i\rho _{i}^{1/2}e^{-i\pi _{i}}~,  \label{Psi}
\end{equation}%
Thus, a point $\psi \in $ $T^{\ast }\mathcal{S}^{+}$ has coordinates 
\begin{equation}
\psi ^{\mu i}=\binom{\psi ^{1i}}{\psi ^{2i}}=\binom{\psi _{i}}{i\psi
_{i}^{\ast }}~,
\end{equation}%
where the index $\mu =1,2$ takes two values (with $\mu ,\nu ,\ldots $ chosen
from the middle of the Greek alphabet).

Since changing the phase $\pi _{i}\rightarrow \pi _{i}+2\pi $ yields the
same point $\psi $ we see that the new $T^{\ast }\mathcal{S}_{\rho }^{+}$ is
a flat $n$-dimensional \textquotedblleft hypercube\textquotedblright\ (its
edges have coordinate length $2\pi $) with the opposite faces identified
(periodic boundary conditions). Thus, the new $T^{\ast }\mathcal{S}_{\rho
}^{+}$ is locally isomorphic to the old $R^{n},$ which makes it a legitimate
choice of cotangent space. (Strictly, $T^{\ast }\mathcal{S}_{\rho }^{+}$ is
a parallelepiped; from (\ref{G+ a}) we see that the lengths of its edges are 
$\ell _{i}=2\pi (2\rho _{i})^{1/2}$ which vanish at the boundaries of the
simplex.)

We can check that the transformation from real $(\rho ,\pi )$ to complex
coordinates $(\psi ,i\psi ^{\ast })$ is canonical, so that 
\begin{equation}
\Omega _{\mu i,\nu j}=%
\begin{bmatrix}
0 & 1 \\ 
-1 & 0%
\end{bmatrix}%
\delta _{ij}~,  \label{sympl form d}
\end{equation}%
retains the same form as (\ref{sympl form b}).

Expressed in $\psi $ coordinates the Hamiltonian flow generated by the
normalization constraint (\ref{momentum shift}) is the familiar phase shift $%
\psi _{i}(\nu )=\psi _{i}(0)e^{i\nu }$. Thus, the gauge symmetry induced by
the constraint $\tilde{N}=0$ is the familiar multiplication by a constant
phase factor.

In $\psi $ coordinates the metric $G$ on $T^{\ast }\mathcal{S}^{+}$ eq.(\ref%
{ePS a}) becomes 
\begin{equation}
\delta \tilde{s}^{2}=-2i\tsum\limits_{i=1}^{n}\,\delta \psi _{i}\delta i\psi
_{i}^{\ast }=G_{\mu i,\nu j}\,\delta \psi ^{\mu i}\delta \psi ^{\nu j}\quad 
\text{where}\quad G_{\mu i,\nu j}=-i\delta _{ij}%
\begin{bmatrix}
0 & 1 \\ 
1 & 0%
\end{bmatrix}%
~.  \label{metric Psi a}
\end{equation}%
Finally, using the inverse $G^{\mu i,\lambda k}$ to raise the first index of 
$\Omega _{\lambda k,\nu j}$ gives the $\psi $ components of the tensor $J$, 
\begin{equation}
J^{\mu i}{}_{\nu j}=-G^{\mu i,\lambda k}\Omega _{\lambda k,\nu j}=%
\begin{bmatrix}
i & 0 \\ 
0 & -i%
\end{bmatrix}%
\delta _{ij}~.  \label{complex structure J}
\end{equation}

\section{Hamilton-Killing flows}

\label{HK flows}We have studied those Hamiltonian flows $\bar{K}$ that, in
addition to preserving the symplectic form, are generated by a gauge
invariant $\tilde{K}$ in order to preserve the normalization constraint $%
\tilde{N}$. Our next goal is to find those flows that also happen to
preserve the metric $G$ of $T^{\ast }\mathcal{S}^{+}$, that is, we want $%
\bar{K}$ to be a Killing vector. The vector field $\bar{K}$ is determined by
the Killing equation \cite{Schutz 1980}, $\pounds _{K}G=0$, or 
\begin{equation}
(\pounds _{K}G)_{\mu i,\nu j}=K^{\lambda k}\partial _{\lambda k}G_{\mu i,\nu
j}+G_{\lambda k,\nu j}\partial _{\mu i}K^{\lambda k}+G_{\mu i,\lambda
k}\partial _{\nu j}K^{\lambda k}=0~.
\end{equation}%
Since eq.(\ref{metric Psi a}) gives $\partial _{\lambda k}G_{\mu i,\nu j}=0$%
, the Killing equation simplifies to 
\begin{equation}
(\pounds _{K}G)_{\mu i,\nu j}=-i%
\begin{bmatrix}
\frac{\partial K^{2j}}{\partial \psi _{i}}+\frac{\partial K^{2i}}{\partial
\psi _{j}}~; & \frac{\partial K^{1j}}{\partial \psi _{i}}+\frac{\partial
K^{2i}}{\partial i\psi _{j}^{\ast }} \\ 
\frac{\partial K^{2j}}{\partial i\psi _{i}^{\ast }}+\frac{\partial K^{1i}}{%
\partial \psi _{j}}~; & \frac{\partial K^{1j}}{\partial i\psi _{i}^{\ast }}+%
\frac{\partial K^{1i}}{\partial i\psi _{j}^{\ast }}%
\end{bmatrix}%
=0~,  \label{K flow a}
\end{equation}%
\noindent where $\partial /\partial i\psi _{i}^{\ast }\overset{\text{def}}{=}%
-i\partial /\partial \psi _{i}^{\ast }$. If we further require that $\bar{K}$
be a Hamiltonian flow, $\pounds _{K}\Omega =0$, then $K^{\mu i}$ satisfies
Hamilton's equations, 
\begin{equation}
K^{1i}=\frac{\partial \tilde{K}}{\partial i\psi _{i}^{\ast }}\quad \text{and}%
\quad K^{2i}=-\frac{\partial \tilde{K}}{\partial \psi _{i}}~.
\label{H flow a}
\end{equation}%
Substituting into (\ref{K flow a}) we find 
\begin{equation}
\frac{\partial ^{2}\tilde{K}}{\partial \psi _{i}\partial \psi _{j}}=0\quad 
\text{and}\quad \frac{\partial ^{2}\tilde{K}}{\partial \psi _{i}^{\ast
}\partial \psi _{j}^{\ast }}=0~.
\end{equation}%
Therefore, in order to generate a flow that preserves both $G$ and $\Omega $
the function $\tilde{K}(\psi ,\psi ^{\ast })$ must be \emph{linear} in both $%
\psi $ and $\psi ^{\ast }$, 
\begin{equation}
\tilde{K}(\psi ,\psi ^{\ast })=\tsum\limits_{i,j=1}^{n}\psi _{i}^{\ast }\hat{%
K}_{ij}\psi _{j}+\tsum\limits_{i=1}^{n}\left( \psi _{i}^{\ast }\hat{L}_{i}+%
\hat{M}_{i}\psi _{i}\right) +\limfunc{const}~,
\end{equation}%
where the kernels $\hat{K}_{ij}$, $\hat{L}_{i}$, and $\hat{M}_{i}$ are
independent of $\psi $ and $\psi ^{\ast }$. Imposing that the flow preserves
the normalization constraint $\tilde{N}=\limfunc{const}$, eq.(\ref{N
conservation}), implies that $\tilde{K}$ must be invariant under the phase
shift $\psi \rightarrow \psi e^{i\nu }$. Therefore, $\hat{L}_{i}=\hat{M}%
_{i}=0$ and we conclude that 
\begin{equation}
\tilde{K}(\psi ,\psi ^{\ast })=\tsum\limits_{i,j=1}^{n}\psi _{i}^{\ast }\hat{%
K}_{ij}\psi _{j}+\limfunc{const}~.  \label{bilinear hamiltonian}
\end{equation}%
The corresponding HK flow is given by Hamilton's equations, 
\begin{align}
\frac{d\psi _{i}}{d\lambda }& =K^{1i}=\frac{\partial \tilde{K}}{\partial
i\psi _{i}^{\ast }}=\frac{1}{i}\tsum\limits_{j=1}^{n}\,\hat{K}_{ij}\psi
_{j}~,  \label{HK flow 1} \\
\frac{di\psi _{i}^{\ast }}{d\lambda }& =K^{2i}=-\frac{\partial \tilde{K}}{%
\partial \psi _{i}}=-\tsum\limits_{j=1}^{n}\,\psi _{j}^{\ast }\hat{K}_{ji}~.
\label{HK flow 2}
\end{align}%
The constant in (\ref{bilinear hamiltonian}) can be dropped because it has
no effect on the flow. Taking the complex conjugate of (\ref{HK flow 1}) and
comparing with (\ref{HK flow 2}), shows that the kernel $\hat{K}_{ij}$ is
Hermitian, and that the corresponding Hamiltonian functionals $\tilde{K}$
are real, 
\begin{equation}
\hat{K}_{ij}^{\ast }=\hat{K}_{ji}\quad \text{and}\quad \tilde{K}(\psi ,\psi
^{\ast })^{\ast }=\tilde{K}(\psi ,\psi ^{\ast })~.  \label{V selfadjoint}
\end{equation}

To summarize: \emph{the preservation of the symplectic structure, the metric
structure, and the normalization constraint leads to Hamiltonian functions }$%
\tilde{K}$\emph{\ that are bilinear in }$\psi $\emph{\ and }$\psi ^{\ast }$,
eq.(\ref{bilinear hamiltonian}). This is the main result of this paper. To
appreciate its significance once again we adopt a more suggestive notation:
the flow generated by the Hamiltonian function 
\begin{equation}
\tilde{H}(\psi ,\psi ^{\ast })=\tsum\limits_{i,j=1}^{n}\psi _{i}^{\ast }\hat{%
H}_{ij}\psi _{j}\quad \text{is}\quad \frac{d\psi _{i}}{d\tau }=\{\psi _{i},%
\tilde{H}\}\quad \text{or}\quad i\frac{d\psi _{i}}{d\tau }%
=\tsum\limits_{j=1}^{n}\,\hat{H}_{ij}\psi _{j}~,  \label{Sch eq}
\end{equation}%
which is recognized as the Schr\"{o}dinger equation. Beyond being Hermitian,
the actual form of the kernel $\hat{H}_{ij}$ remains undetermined.\ 

The central feature of Hamilton's equations (\ref{HK flow 1}) or of the Schr%
\"{o}dinger equation (\ref{Sch eq}) is that they are linear. Given two
solutions $\psi ^{(1)}$ and $\psi ^{(2)}$ and arbitrary constants $c_{1}$
and $c_{2}$, the linear combination $\psi ^{(3)}=c_{1}\psi ^{(1)}+c_{2}\psi
^{(2)}$ is a solution too and this is extremely useful in calculations.
Unfortunately, this is an HK flow on the embedding space $T^{\ast }\mathcal{S%
}^{+}$ and when the flow is projected onto the e-phase space $T^{\ast }%
\mathcal{S}$ the linearity is severely restricted by normalization. If $\psi
^{(1)}$ and $\psi ^{(2)}$ are normalized points on $T^{\ast }\mathcal{S}$
the superposition $\psi ^{(3)}$ will not in general be a normalized point on 
$T^{\ast }\mathcal{S}$ unless the constants $c_{1}$ and $c_{2}$ are chosen
appropriately. Furthermore, the states $\psi ^{\prime (1)}=\psi
^{(1)}e^{i\nu _{1}}$ and $\psi ^{\prime (2)}=\psi ^{(2)}e^{i\nu _{2}}$ are
supposed to be \textquotedblleft physically\textquotedblright\ equivalent to
the original $\psi ^{(1)}$ and $\psi ^{(2)}$, but in general the
superposition $\psi ^{\prime (3)}=c_{1}\psi ^{\prime (1)}+c_{2}\psi ^{\prime
(2)}$ is not equivalent to $\psi ^{(3)}$. In other words, \emph{the
mathematical linearity of (\ref{HK flow 1}) or (\ref{Sch eq}) does not
extend to a full blown Superposition Principle for physically equivalent
states}. On the other hand, any point $\psi $ deserves to be called a
\textquotedblleft state\textquotedblright\ in the limited sense that it may
serve as the initial condition for a curve in $T^{\ast }\mathcal{S}^{+}$.
Since given two states $\psi ^{(1)}$ and $\psi ^{(2)}$ their superposition $%
\psi ^{(3)}$ is a state too, we see that the set of states $\{\psi \}$ forms
a linear vector space. This is a structure that turns out to be very useful.

\section{Hilbert space}

We just saw that the possible initial conditions for an HK flow, the \emph{%
points} $\psi $ of $T^{\ast }\mathcal{S}^{+}$, form a linear vector space.
To take full advantage of linearity we would like to endow this vector space
with the additional structure of an inner product and turn it into a Hilbert
space --- a term which we use loosely to describe any complex vector space
with a Hermitian inner product. The metric tensor $G$, eq.(\ref{metric Psi a}%
), and the symplectic form $\Omega $, eq.(\ref{sympl form d}), are supposed
to act on \emph{vectors} $d/d\lambda $; their action on the \emph{points} $%
\psi $ or $(\rho ,\pi )$ is not defined. The choice of inner product for the
points $\psi $ is, however, natural in the sense that the necessary
ingredients, $G$ and $\Omega $, are already available.

We adopt the familiar Dirac notation to represent the states $\psi $ as
vectors $|\psi \rangle $. The inner product $\langle \psi |\phi \rangle $ is
defined in terms of the tensors $G$ and $\Omega $, 
\begin{equation}
\langle \psi |\phi \rangle =\frac{1}{2}\left( G_{\mu i,\nu j}+\alpha \Omega
_{\mu i,\nu j}\right) \psi ^{\mu i}\phi ^{\nu j}~,  \label{inner prod a}
\end{equation}%
where $\alpha $ is a constant and, to follow convention, an overall constant
has been set to $1/2$. Using eq.(\ref{sympl form d}) and (\ref{metric Psi a}%
) we get 
\begin{equation}
\langle \psi |\phi \rangle =\frac{1}{2}\left( \psi _{i},i\psi _{i}^{\ast
}\right) \left( G+\alpha \Omega \right) \binom{\phi _{j}}{i\phi _{j}^{\ast }}%
=\frac{1}{2}\tsum\limits_{i=1}^{n}\left( (1-i\alpha )\psi _{i}^{\ast }\phi
_{i}+(1+i\alpha )\phi _{i}^{\ast }\psi _{i}\right) ~.  \label{inner prod b}
\end{equation}%
To fix $\alpha $ we impose that $\langle \psi |\phi \rangle ^{\ast }=\langle
\phi |\psi \rangle $ which implies that $\alpha =$ $\pm i$. In order to
comply with the standard convention that the inner product $\langle \psi
|\phi \rangle $ be linear in the second factor and anti-linear in the first
factor, we select $\alpha =+i$. The result is the familiar expression for
the positive definite inner product, 
\begin{equation}
\langle \psi |\phi \rangle \overset{\text{def}}{=}\frac{1}{2}\left( G_{\mu
i,\nu j}+i\Omega _{\mu i,\nu j}\right) \psi ^{\mu i}\phi ^{\nu
j}=\tsum\limits_{i=1}^{n}\psi _{i}^{\ast }\phi _{i}~.
\end{equation}%
Here we see that the choice of $1/2$ as the overall constant has led to the
standard relation $\langle \psi |\psi \rangle =|\rho |$. The map between
points and vectors, $\psi \leftrightarrow |\psi \rangle $, is defined by $%
|\psi \rangle =\sum_{i}\,|i\rangle \psi _{i}\ $where $\psi _{i}=\langle
i|\psi \rangle $, and the vectors $\{|i\rangle \}$ form a basis that is
orthogonal and complete.

The bilinear Hamilton function $\tilde{K}(\psi ,\psi ^{\ast })$ with kernel $%
\hat{K}_{ij}$ can now be written as the expected value, $\tilde{K}(\psi
,\psi ^{\ast })=\langle \psi |\hat{K}|\psi \rangle $, of the Hamiltonian
operator $\hat{K}$ with matrix elements $\hat{K}_{ij}=\langle i|\hat{K}%
|j\rangle $. The corresponding HK flows are given by 
\begin{equation}
i\frac{d}{d\lambda }\langle i|\psi \rangle =\langle i|\hat{K}|\psi \rangle
\quad \text{or}\quad i\frac{d}{d\lambda }|\psi \rangle =\hat{K}|\psi \rangle
~,
\end{equation}%
which are described by unitary transformations 
\begin{equation}
|\psi (\lambda )\rangle =\hat{U}_{K}(\lambda )|\psi (0)\rangle \quad \text{%
where}\quad \hat{U}_{K}(\lambda )=\exp (-i\hat{K}\lambda )~.
\end{equation}
Finally, the Poisson bracket of two Hamiltonian functions $\tilde{U}[\psi
,\psi ^{\ast }]$ and $\tilde{V}[\psi ,\psi ^{\ast }]$, can be written in
terms of the commutator of the associated operators, 
\begin{equation}
\{\tilde{U},\tilde{V}\}=-i\langle \psi |[\hat{U},\hat{V}]|\psi \rangle ~.
\end{equation}
Thus the Poisson bracket is the expectation of the commutator. This \emph{%
identity} is much sharper than Dirac's pioneering discovery that the quantum
commutator of two quantum variables is \emph{analogous} to the Poisson
bracket of the corresponding classical variables.

\section{Conclusion}

There have been numerous attempts to derive or construct the mathematical
formalism of quantum mechanics by adapting the symplectic geometry of
classical mechanics. Such phase space methods invariably start from a
classical phase space of positions and momenta $(q^{i},p_{i})$ and through
some series of \textquotedblleft quantization rules\textquotedblright\ posit
a correspondence to self-adjoint operators $(\hat{Q}^{i},\hat{P}_{i})$ which
no longer constitute a phase space. The connection to classical mechanics is
lost. The interpretation of $\hat{Q}^{i}$ and $\hat{P}_{i}$ and even the
answer to the question of what is real or \emph{ontic} and what is \emph{%
epistemic} all become highly controversial. Probabilities play a secondary
role in such formulations.

In this paper we have taken a different starting point that places
probabilities at the forefront. We have discussed special families of curves
--- the Hamilton-Killing flows --- that promise to be useful for the study
of quantum mechanics. We have shown that the HK flows that preserve the
symplectic and the metric structures of the e-phase space reproduce much of
the mathematical formalism of quantum theory. It clarifies how the linearity
of the Schr\"{o}dinger equation, complex numbers, the Born rule $\rho
_{i}=|\psi _{i}|^{2}$ (the Born rule for generic observables is discussed in 
\cite{Johnson Caticha 2011}\cite{Vanslette Caticha 2016}), all follow from
the symplectic and metric structures, while the normalization constraint
leads to the equivalence of states along rays in a Hilbert vector space.

\subparagraph*{Acknowledgments}

I would like to thank M. Abedi, D. Bartolomeo, C. Cafaro, N. Carrara, N.
Caticha, F. X. Costa, S. DiFranzo, S. Ipek, D.T. Johnson, S. Nawaz, P.
Pessoa, M. Reginatto, and K. Vanslette, for valuable discussions and for
their many insights and contributions at various stages of this program.

\end{document}